\documentstyle[preprint,prd,aps,eqsecnum,array,epsf]{revtex}

\tightenlines

\begin{document}

\preprint{DFT-IF-UERJ-99/10}

\title{High Temperature Resummation in the Linear $\delta$-Expansion}

\author{Marcus B. Pinto$^1\;$\thanks{E-mail: fsc1mep@fsc.ufsc.br }
and Rudnei O. Ramos$^2\;$\thanks{E-mail: rudnei@dft.if.uerj.br}} 

\address{
{\it $^1\;$ Departamento de F\'{\i}sica,}
{\it  Universidade Federal de Santa Catarina,}\\
{\it 88.040-900 Florian\'{o}polis, SC, Brazil}\\
{\it $^2\;$ Departamento de F\'{\i}sica Te\'orica - Inst. de F\'{\i}sica,}
{\it Universidade do Estado do Rio de Janeiro, }\\
{\it 20.550-013 Rio de Janeiro, RJ, Brazil}}

\maketitle

\thispagestyle{empty}

\begin{abstract}

The optimized linear $\delta$-expansion is applied to the $\lambda
\phi^4$ theory at high temperature. Using the imaginary time formalism
the thermal mass is evaluated perturbatively up to order $\delta^2$. A
variational procedure associated with the method generates
nonperturbative results which are used to obtain the critical
temperature for the phase transition. Our results are compared with 
the ones given by propagator dressing methods. 

\vspace{0.34cm}
\noindent
PACS number(s): 11.15.Tk, 12.38.Lg

\end{abstract}

\newpage

\setcounter{page}{1}

\section{Introduction}

The breakdown of perturbation expansion in high temperature
quantum field theory is a well known problem \cite{pert1,pert2} whose solution
is still a matter of study and discussion today, with different authors
using different methods
\cite{Espinosa,sato,nami,Camelia,Banerjee,Parwani,Boyd,sato2,Hatsuda}. 
High temperature
perturbation expansion breaks down due to the appearance of infrared
divergences close to critical temperatures (in field theories displaying a
second order phase transition or a weakly first order transition
\cite{GR,beyond}), or for massless field theories, like QCD. In
particular, there are parameter regimes where conventional perturbation
schemes become unreliable at high temperature when powers of the
coupling constants become surmounted by powers of the temperature. 

In general, these problems are treated with resummation techniques which
try to account, in a self-consistent way, for the leading contributions
in the infrared region. Among these schemes are the popular daisy and
super-daisy schemes \cite{Espinosa,sato}, composite operator method
\cite{Camelia} and field propagator dressing methods
\cite{Banerjee,Parwani}. Some of these resummation methods have
been compared in Ref. \cite{Boyd}, where their difficulties and possible
caveats have also been discussed. The majority of approaches used within
this subject have a potential drawback concerning the achievement of
self-consistency as higher order diagrams are resummed. This happens
because the dressed propagator changes order by order. 
Therefore, special care
must be taken when selecting the correct order in the coupling constants.
Another problem associated with some of these methods is related to the 
implementation of renormalization, as discussed in 
Ref. \cite{Hatsuda}.

In this paper, we apply the optimized linear $\delta$-expansion
\cite{linear,du} (for earlier references see, {\it e.g.}, \cite{seznec})
to the $\lambda \phi^4$ theory obtaining the thermal
mass to second order in the perturbative parameter $\delta$. Our results
show that the use of a proper optimization scheme is equivalent to solve
the gap equation for the thermal mass, where leading and higher order
infrared regularizing contributions are nonperturbatively taken into
account. An advantage of the linear $\delta$-expansion is that the same
simple propagator is used in the evaluation of any diagram, avoiding the
potential bookkeeping problems mentioned above. This makes the method
particularly useful and simple to use in different applications,
including the study of nonperturbative high temperature effects.

This work is organized as follows. In Sec. II we briefly describe the
linear $\delta$-expansion technique and then use it to evaluate the
thermal mass up to order $\delta^2$ in the $3+1d$ $\lambda \phi^4$
theory. Details of the renormalization up to this order are given 
in Sec. III where we also discuss renormalization at higher orders
in $\delta$. We 
show that it does not present any additional difficulty when compared to
renormalization in the usual perturbative or loop expansions. 
In Sec. IV we present our
results for the thermal mass, including the critical temperature for the
phase transition, and compare them with other results found in the
literature. In Sec. V concluding remarks are given.

\section{The Linear $\delta$-Expansion Applied to the
Evaluation of the Thermal Mass in the $\lambda \phi^4$ Theory}

\subsection{The Linear $\delta$-Expansion}

The optimized linear $\delta$-expansion is an
alternative nonperturbative approximation which has been
successfully used in  a plethora of different problems in  particle theory 
\cite{du,okotem,landau,njlft}, quantum mechanics \cite {ian,guida} 
statistical physics \cite {alan}, nuclear matter \cite {gas} and lattice 
field theory \cite{evans}. One advantage of this method is that
the selection and evaluation (including renormalization) of Feynman
diagrams are done exactly as in ordinary perturbation theory using a
very simple modified propagator which depends on an arbitrary mass
parameter. Nonperturbative results are then obtained by fixing this
parameter. An interesting result obtained with this method in the finite
temperature domain is given in Ref.~\cite {landau} where the
critical temperature value for the Gross-Neveu model in 1+1 dimensions 
nicely converges, order by order, towards the exact result set by Landau's 
theorem ($T_c=0$) \footnote{Consistent results in the $\lambda \phi^4$ theory 
at finite 
temperature have also been obtained with  a variant of the linear 
$\delta$-expansion, the nonlinear $\delta$-expansion \cite{rebhan,rudnei}.
However, beyond first
order, this latter version presents cumbersome technical problems associated with the evaluation of graphs.}.

The standard application of the linear
$\delta$-expansion to a theory described by a Lagrangian density 
${\cal L}$ starts with an interpolation defined by 
\begin{equation}
{\cal L}^{\delta} = (1-\delta){\cal L}_0(\eta) + \delta {\cal L} = 
{\cal L}_0(\eta) + \delta [{\cal L}-{\cal L}_0(\eta)],
\label{int}
\end{equation}

\noindent
where ${\cal L}_0(\eta)$ is the Lagrangian density of a solvable theory
which can contain an arbitrary mass parameter ($\eta$). The Lagrangian
density ${\cal L}^{\delta}$ interpolates between the solvable ${\cal
L}_0(\eta)$ (when $\delta=0$) and the original ${\cal L}$ (when
$\delta=1$). In this work we consider the $\lambda \phi^4$ model
described by

\begin{equation}
{\cal L} = \frac{1}{2} (\partial_{\mu}\phi)^2 - \frac{1}{2} m^2   
\phi^2 - \frac {\lambda}{4!}
\phi^4 + {\cal L}_{\rm ct}\;,
\end{equation}
where
\begin{equation}
{\cal L}_{\rm ct}=  A \frac{1}{2} (\partial_{\mu}\phi)^2 - 
B \frac{1}{2} m^2 \phi^2 - 
\frac {\lambda}{4!} C \phi^4  \;\;
\end{equation}
represents the counterterms needed to render the 
model finite. Note that ${\cal L}_{\rm ct}$ requires an extra piece if one attempts to evaluate the thermal effective potential \cite{Hatsuda}, which is not the case here.
Choosing
\begin{equation}
{\cal L}_0(\eta) = \frac{1}{2} (\partial_{\mu}\phi)^2 - 
\frac{1}{2} m^2 \phi^2 - 
\frac{1}{2}  \eta^2 \phi^2 
\end{equation}
and following the general prescription one can write
\begin{equation}
{\cal L}^{\delta} = \frac{1}{2} (\partial_{\mu}\phi)^2 - 
\frac{1}{2} m^2 \phi^2 - \frac {\delta
\lambda}{4!} \phi^4 - \frac{1}{2} (1- \delta) \eta^2 \phi^2 + 
{\cal L}_{\rm ct}^{\delta}\;,
\end{equation}
or
\begin{equation}
{\cal L}^{\delta} = \frac{1}{2} (\partial_{\mu}\phi)^2 - 
\frac{1}{2} \Omega^2 \phi^2 - \frac
{\delta \lambda}{4!} \phi^4 + \delta \frac{1}{2} \eta^2 \phi^2 + 
{\cal L}_{\rm ct}^{\delta}\;,
\end{equation}
where $\Omega^2=m^2+\eta^2$ and
\begin{equation}
{\cal L}_{\rm ct}^{\delta}=A^{\delta}\frac{1}{2} (\partial_{\mu}\phi)^2 
- \frac{1}{2} \Omega^2 
B^{\delta}\phi^2 - \frac{\delta \lambda}{4!} C^{\delta}\phi^4 + 
\delta {\frac{1}{2}} \eta^2 B^{\delta}\phi^2\;.
\end{equation}

\noindent
One should note that the $\delta$-expansion interpolation introduces
only ``new" quadratic terms not altering the renormalizability of the
original theory. That is, the counterterms contained in ${\cal
L}^{\delta}_{\rm ct}$, as well as in the original ${\cal L}_{\rm ct}$,
have the same polynomial structure.

The general way the method works becomes clear by looking at the Feynman
rules generated by ${\cal L}^{\delta}$. First, the original $\phi^4$
vertex has its original Feynman rule $-i \lambda$ modified to $-i\delta
\lambda$. This minor modification is just a reminder that one is really
expanding in orders of the artificial parameter $\delta$. Most
importantly, let us look at the modifications implied by the addition of
the arbitrary quadratic part. The original bare propagator, 

\begin{equation}
S(k)= i(k^2-m^2 +i\epsilon)^{-1}\;, 
\end{equation}

\noindent
becomes
\begin{equation}
S(k)= i(k^2-\Omega^2 +i\epsilon)^{-1}=
{i \over {k^2 - m^2 + i\epsilon  }}\left[ 1 - 
{\frac{i}{k^2-m^2 +i\epsilon} (-i\eta^2)}
\right ]^{-1}\,,
\label{prop}
\end{equation}

\noindent
indicating that the term proportional to $ \eta^2 \phi^2$ contained in
${\cal L}_0$ is entering the theory in a nonperturbative way. On the
other hand, the piece proportional to $\delta\eta^2 \phi^2$ is only
being treated perturbatively as a quadratic vertex (of weight $i \delta
\eta^2$). Since only an infinite order calculation would be able to
compensate for the infinite number of ($-i\eta^2$) insertions contained
in Eq.~(\ref {prop}), one always ends up with a $\eta$ dependence in any
quantity calculated to finite order in $\delta$. Then, at the end of the
calculation one sets $\delta=1$ (the value at which the original theory
is retrieved) and fixes $\eta$ with the variational procedure known as
the Principle of Minimal Sensitivity (PMS) \cite{pms} \footnote{The 
third refence in \cite{ian} discusses alternative conditions for fixing 
$\eta$.}

\begin{equation}
\frac{\partial P(\eta)}{\partial \eta} |_{\bar \eta} =0\;,
\end{equation}
where $P$ represents a physical quantity calculated 
{\it perturbatively} in 
powers of $\delta$.

This optimization procedure, together with the convergence problem, has 
been discussed in detail for simple cases in low dimensions in 
Refs.~\cite {ian} and \cite {guida} where possible implications to  
more realistic theories have also been investigated. Both references 
provide proofs of convergence. Using the anharmonic oscillator, 
Bellet {\it et al.} \cite{phil} have also studied the convergence of an 
alternative version of the linear $\delta$-expansion. Their method has 
been later extended to the Gross-Neveu model where the optimization 
procedure was studied in conjunction with the renormalization group 
\cite{jloic}. The convergence of the $\delta$-expansion in
quantum field theory is still a subject deserving further investigation. 
In principle, it seems that if the convergence
proof conditions for the anharmonic oscillator could be extended
to quantum field theory in $d < 4$ \cite{guida},
one could pursue a similar investigation for the $\lambda \phi^4$
theory in $3+1 \; d$ at finite temperature.
This is due to the fact that at very high temperatures this model 
gets dimensionally reduced 
to an effective $3d$ theory for the zero Matsubara field modes 
\cite{reduction}. However, we shall not pursue this discussion.

As far as renormalization is concerned it is important to note that 
in general, as a result of the optimization procedure, the arbitrary
$\eta$ turns out to be a function of the original model parameters,
scales introduced through regularization as well as external parameters
such as the temperature and/or density. Therefore, in order to get
physically acceptable results the optimization procedure must be carried
out after all divergences have been eliminated. The renormalization problem, 
in the large-$N$ limit, has been addressed in Ref. \cite {moshe}. The way 
renormalization will be carried out here is well discussed in 
Ref. \cite{Hatsuda}.

\subsection{The Evaluation of the Thermal Mass at order $\delta^2$}

We can now start our evaluation of the thermal mass, defined by
\begin{equation}
M_T^2= \Omega^2 + \Sigma^{\delta}_T(p)\;,
\end{equation}
where $\Sigma^{\delta}_T(p)$ is the thermal self-energy. 
At lowest order (first
order in $\delta$) the relevant contributions, which are momentum
independent, are shown in {}Fig. 1 and given by

\vspace{0.5cm}
\begin{equation}
\Sigma_T^{\delta^1} (p) =
- \delta \eta^2 + \delta \frac {\lambda}{2}  \int_T
\frac{d^4k}{(2\pi)^4} \frac {i}{k^2-\Omega^2+i\epsilon}\;.
\label{delta1}
\end{equation}

\noindent
The temperature dependence can be readily obtained by using the 
imaginary time formalism prescription (see, {\it e.g.} \cite{pert1})

\begin{equation}
p_0 \rightarrow i \omega_n \;\;, \;\;\;\;\;\; \int_T 
\frac{d^4k}{(2\pi)^4} \rightarrow i T \sum \int\frac{d^3{\bf
k}}{(2\pi)^3}\;.
\end{equation}
Then, the self-energy becomes

\begin{equation}
\Sigma^{\delta^1}_T(p) = - \delta \eta^2 + \delta  T \frac {\lambda}{2} 
\sum \int 
\frac{d^3 {\bf
k}}{(2\pi)^3} \frac {1}{\omega_n^2+E^2}\;,
\end{equation}
where $E^2={\bf k}^2+\Omega^2$. Summing over Matsubara's frequencies one 
gets
\begin{equation}
\Sigma^{\delta^1}_T(p)= - \delta \eta^2 + \delta \frac{\lambda}{2}  
\int \frac{d^3 {\bf
k}}{(2\pi)^3} \left \{ \frac {1}{2 E} - \frac {1}{E[
1 -\exp(E/T)]}\right \}\;.
\end{equation}
Then, using dimensional regularization \cite{ra} one obtains the thermal mass
\begin{equation}
M_{T}^2 = \Omega^2 - \delta \eta^2 + 
\delta\frac{\lambda}{32 \pi^2}\Omega^2
\left [ - \frac {1}{\epsilon} + 
\ln \left (\frac{\Omega^2}{4 \pi \mu^2} \right)+
\gamma_E-1
\right] + \delta \lambda T^2 h\left(\frac{\Omega}{T}\right)\;,
\label{mter1}
\end{equation}
where $\mu$ is a mass scale introduced by dimensional regularization and  
\begin{equation}
h(y) = \frac {1}{4 \pi^2}  \int_0^{\infty} dx \frac
{x^2}{[x^2+y^2]^{\frac{1}{2}}[ \exp(x^2+y^2)^{\frac{1}{2}}-1] }\;,
\end{equation}
where $x=k/T$.
Note that the temperature independent term diverges and must be 
renormalized. In this paper we chose the Minimal Subtraction (MS) 
scheme where the counterterms eliminate the poles only. At this order 
the only divergence is
\begin{equation}
\Sigma^{\delta^1}_{\rm div}(\Omega^2) = - \delta \frac{  
\lambda \Omega^2}{32 \pi^2 \epsilon}\;,
\label{div1}
\end{equation}
which is easily eliminated by the ${\cal O}(\delta)$ mass 
counterterm 
\begin{equation}
\Sigma^{\delta^1}_{\rm ct}(\Omega^2) = B^{\delta^1} \Omega^2 = \left( 
\delta \frac{  
\lambda }{32 \pi^2 \epsilon} \right ) \Omega^2 \;.
\label{ct1}
\end{equation}
By looking at Eq. (\ref{mter1}) one can see that the terms proportional 
to $\delta \lambda$ represent exactly the same diagram which appears at 
${\cal O}(\lambda)$ in ordinary $\lambda \phi^4$ theory, excepted that we 
now have $\Omega^2$ instead of $m^2$ and $\delta \lambda$ instead of 
$\lambda$. Therefore, it is not surprising that to this order the 
renormalization procedure implied by the interpolated theory is 
identical to the procedure implied by the original theory at 
${\cal O}(\lambda)$. 

Let us now analyze the temperature dependent integral which is expressed, 
in the
high temperature limit ($\Omega/T \ll 1$), as \cite{Kapusta}
\begin{equation}
h(y)=
\frac {1}{24} - \frac{1}{8\pi} y
- \frac{1}{16\pi^2} y^2 \left[ \ln \left ( \frac
{y}{
4\pi} \right ) + \gamma_E - \frac{1}{2} \right ] + \ldots
\end{equation}
In principle, since $\eta$ is arbitrary, one could be reluctant in 
taking the limit $\Omega/T \ll 1$. Therefore, to be sure that the 
PMS can be safely applied to the thermal mass in the high temperature 
limit, we have performed numerical calculations using both forms for 
the integral $h(y)$ finding that the optimization results do not lead
to any significant numerical changes.
Then, taking the integral $h(y)$ in the high temperature limit, 
one obtains the ${\cal O}(\delta)$ thermal mass: 

\begin{equation}
M_T^2= \Omega^2 - \delta \eta^2 +\delta  \frac{\lambda T^2}{24}-\delta 
\frac{\lambda T\Omega}{8\pi} +\delta 
\frac{ \lambda \Omega^2 }{32\pi^2}\left [
\ln\left(\frac{4 \pi T^2}{\mu^2}\right ) - \gamma_E \right ] \;.
\end{equation}

At ${\cal O}(\delta^2)$ the self-energy receives contributions from
momentum independent as well as momentum dependent diagrams. 
At this order there are five diagrams contributing, which are 
shown in {}Fig. 2. Let us
first consider the momentum independent diagram given by the
first diagram in {}Fig. 2, which we call $\Sigma_1^{\delta^2}$.
In the high temperature approximation it is given by

\begin{equation}
\Sigma_1^{\delta^2} \simeq
\delta^2 \frac{\lambda T \eta^2}{16\pi \Omega} - \delta^2 
\frac{ \lambda \eta^2}{32\pi^2}
\left [ - \frac {1}{\epsilon} + 
\ln\left(\frac{ 4 \pi T^2}{\mu^2}\right )  - 
\gamma_E \right
] \;\;\;.
\label{div2}
\end{equation}

\noindent
As will be shortly seen, this contribution can be rendered finite using a mass type counterterm 
contained in ${\cal L}^{\delta}_{\rm ct}$ which is tailored to account for 
divergences arising from the extra quadratic vertex introduced during 
the interpolation process. 

Considering the ${\cal O}(\delta)$ mass counterterm used 
to eliminate 
the divergence in $\Sigma_{\rm div}^{\delta^1}$ (see Eq. (\ref{ct1})), 
one is able to
build a one loop ${\cal O}(\delta^2)$ diagram whose contribution is
given by the second diagram in {}Fig. 2. In the high temperature
approximation one obtains

\begin{eqnarray}
\Sigma_2^{\delta^2} &\simeq&
-\delta^2 \frac { \lambda^2 \Omega^2}{(32 \pi^2)^2 \epsilon^2} +
\delta^2 \frac {\lambda^2}{32 \pi^2 \epsilon} \left \{ - 
\frac {T \Omega}{ 16 \pi} 
+\frac {\Omega^2}{32 \pi^2} \left [ \ln \left (\frac {4 \pi T^2}{\mu^2}
\right ) - \gamma_E  \right ] \right \}\nonumber \\
&-&\delta^2 \frac { \lambda^2 \Omega^2}{2 (32 \pi^2)^2} \left\{ 
\left[\ln \left (
\frac{\Omega^2}{4\pi \mu^2} \right ) + \gamma_E \right]^2 +
\frac{\pi^2}{6} \right\}\;.
\label{div4}
\end{eqnarray}

\noindent
Next, one considers the vertex counterterm, whose Feynman rule, 
$-3 i \delta^2 \lambda^2/(32 \pi^2 \epsilon)$,
can be obtained, as in Ref.~\cite {ra}, by evaluating the order-$\delta^2$ 
contribution to the four point function. The one loop graph evaluated 
with this counterterm, given by the third diagram in {}Fig. 2, gives

\begin{eqnarray}
\Sigma_3^{\delta^2} &\simeq& 
-\delta^2 \frac {3 \lambda^2 \Omega^2}{(32 \pi^2)^2 \epsilon^2} +
\delta^2 \frac {3\lambda^2}{32 \pi^2 \epsilon} \left \{ \frac {T^2}{24} - 
\frac {T \Omega}{ 8 \pi} 
+\frac {\Omega^2}{32 \pi^2} \left [ \ln \left (\frac {4 \pi T^2}{\mu^2}
\right ) - \gamma_E \right ] \right \}\nonumber \\
&-&\delta^2 \frac { 3\lambda^2 \Omega^2}{2 (32 \pi^2)^2}\left\{ 
\left[\ln \left (
\frac{\Omega^2}{4\pi \mu^2} \right ) + \gamma_E -1 \right]^2
+1+\frac{\pi^2}{6} \right\} \;.
\label{div6}
\end{eqnarray}

The next momentum independent contribution is given by the first
two loop diagram shown in {}Fig. 2:

\begin{eqnarray}
\lefteqn{
\Sigma_4^{\delta^2} \simeq \delta^2 \frac {\lambda^2 \Omega^2}{(32\pi^2)^2} 
\frac{1}{\epsilon^2}  -
\delta^2 \frac{\lambda^2}{32\pi^2} \frac {1}{\epsilon} 
\left \{ \frac {T^2}{24} - 
\frac {3 T \Omega}{16 \pi} + \frac {\Omega^2}{16\pi^2}\left [
\ln\left(\frac{4 \pi T^2}{\mu^2}\right ) - \gamma_E \right ] 
\right \}} 
\nonumber \\
&-& \delta^2 \lambda^2 \frac{T^3}{384 \pi \Omega}+\delta^2 
\lambda^2 \frac{T^2}{128\pi^2}
\nonumber \\
&+&\delta^2 \frac{ \lambda^2}{(16\pi)^2}\left\{ \frac {T^2}{3} - 
\frac{3T\Omega}{2\pi} + \frac {\Omega^2}{4 \pi^2}
\left [ \ln\left(\frac{4 \pi T^2}{\mu^2}\right ) - \gamma_E \right ] 
\right\}
\left [ \ln\left(\frac{4 \pi T^2}{\mu^2}\right ) - \gamma_E \right ] 
\nonumber \\
&+&\delta^2 \frac { \lambda^2 \Omega^2}{(32 \pi^2)^2}
\left\{ \left[ 
\ln \left (
\frac{\Omega^2}{4\pi \mu^2} \right ) + \gamma_E - \frac{1}{2}\right]^2
+\frac{3}{4} + \frac{\pi^2}{6} \right\}\;.
\label{div3}
\end{eqnarray}
To render this diagram finite one needs  mass and vertex counterterms 
\cite{ra}. 

The final contribution to the self-energy at ${\cal O}(\delta^2)$ 
comes from the two-loop ``setting sun"
diagram shown by the last term in {}Fig. 2. 
This is a momentum dependent contribution which is given by the real part of

\begin{equation}
\Sigma_5^{\delta^2} =
- \delta^2 \frac{\lambda^2}{6}  \left(G_0
+ G_1 + G_2 \right) \;,
\label{div19}
\end{equation}

\noindent
where $G_0$ is the zero temperature part (in Euclidean time)
of the diagram and $G_1$ and
$G_2$ are the finite temperature ones (with one and two Bose factors,
respectively). ${\rm Re}[ G_0]$ is given by ($d=4-2 \epsilon$):

\begin{equation}
{\rm Re}[ G_0 (p)]  = \mu^{4 \epsilon} \int \frac{d^d k}{(2 \pi)^d} \int
\frac{d^d q}{(2 \pi)^d} \frac{1}{k^2 + \Omega^2} \frac{1}{q^2 +
\Omega^2} \frac{1}{(p-k-q)^2 + \Omega^2} \;.
\end{equation}

\noindent
This contribution has been evaluated in details in Ref. \cite{twoloop}
where  the quoted result is:

\begin{eqnarray}
\lefteqn{{\rm Re}[ G_0 (p)]  = \frac{\mu^{4 \epsilon}}{(2 \pi)^{2d}}   
\frac{\pi^{d+1}
(\Omega^2)^{d-3}}{\sin \pi (\frac{d}{2}-2)} \sum_{k,n=0}^\infty (-1)^n
\left(\frac{p}{\Omega}\right)^{2 n} \frac{1}{n ! \Gamma(\frac{d}{2}+n)} 
\times}
\nonumber \\
& & \times \left[ \frac{\Gamma(\frac{d}{2}+k) B(1+k,1+k) 
\Gamma(2-\frac{d}{2}+k)}
{(k-n)! \Gamma(\frac{d}{2}+k-n)} -  
\frac{\Gamma(2+k) B(3-\frac{d}{2}+k,3-\frac{d}{2}+k)
\Gamma(4-d+k)}{(k-n+1)! \Gamma(3+k-n-\frac{d}{2})} \right] \nonumber \\
& & + \frac{\mu^{4 \epsilon}}{(2 \pi)^{2d}} \pi^d (\Omega^2)^{d-3} 
\Gamma\left(\frac{d}{2}-1
\right) 
\sum_{n=0}^\infty (-1)^n
\frac{\Gamma(3-d+n)}{\Gamma(\frac{d}{2}+n)} 
\left(\frac{p}{\Omega} \right)^{2 n}
B\left(2-\frac{d}{2}+n,2-\frac{d}{2}+n\right) \;,
\end{eqnarray}

\noindent
where $B(x,y)$ is the Beta function:
$B(x,y) = [\Gamma(x) \Gamma(y)]/\Gamma(x+y)$. In what follows we
evaluate the self-energy on-shell ($\vec p= {\bf 0}$, $p_0 = -i \Omega$).
{}For $\epsilon \to 0$, we obtain the following result for the above
expression:

\begin{eqnarray}
\lefteqn {-\frac{\delta^2\lambda^2}{6} {\rm Re} 
[G_0 (- i\Omega,{\bf 0})]  =
\frac{\delta^2 \lambda^2 \Omega^2}{4 (4 \pi)^4} 
\left[  \frac{1}{\epsilon^2} +
\frac{3-2 \gamma_E}{\epsilon} - \frac{2}{\epsilon} 
\ln\left( \frac{
\Omega^2}{4 \pi \mu^2} \right) \right] +
\frac{\delta^2 \lambda^2 p^2}{4 (4 \pi)^4}  
\frac{1}{6 \epsilon} }\nonumber \\
&& + \frac{\delta^2 \lambda^2 \Omega^2}{2 (4 \pi)^4} 
\left[ \ln^2 \left( \frac{
\Omega^2}{4 \pi \mu^2} \right) + \left(2 \gamma_E -\frac{17}{6}\right) 
\ln\left(
\frac{
\Omega^2}{4 \pi \mu^2} \right) + \gamma^2_E -\frac{17 \gamma_E}{6} +
3.5140 \right]\;,
\label{div7}
\end{eqnarray}

\noindent
where we purposefully left the momentum dependence in the relevant
divergent term
to make explicit the need for a wave-function renormalization counterterm.

The finite temperature contributions $G_1$ and $G_2$ are given, as
in Ref.~ \cite {Parwani}, by
\begin{equation}
-\delta^2 \frac{\lambda^2}{6} {\rm Re} [G_1 (-i \Omega,{\bf 0})]  = {}F_0 + 
{}F_1 + {}F_2 \;,
\end{equation}

\noindent
where

\begin{equation}
{}F_0 = - \delta^2 \frac{\lambda^2 T^2}{ (4\pi)^2} \frac{1}{\epsilon} 
h\left(\frac{\Omega}{T}\right) 
\;,
\label{div8}
\end{equation}

\begin{equation}
{}F_1 = - \delta^2\frac{\lambda^2 T^2}{2 (4 \pi)^2} 
h\left(\frac{\Omega}{T}\right) \left [ - \ln\left(
\frac{
\Omega^2}{4 \pi \mu^2} \right) +2 - \gamma_E \right ]
\end{equation}

\noindent
and

\begin{equation}
{}F_2 = - \delta^2\frac{\lambda^2}{8 (2 \pi)^4} \int_0^\infty dk \frac{k}{E_k
\left(e^{\beta E_k} -1 \right)} \int_0^\infty \frac{dq}{E_q} \left[ q
\ln \left| \frac{X_+}{X_-} \right| - 4 k \right] \;,
\end{equation}

\begin{equation}
X_\pm = \left[ \Omega^2 - \left(E_k+E_q+E_{k\pm q} \right)^2\right]
\left[ \Omega^2 - \left(-E_k+E_q+E_{k\pm q} \right)^2\right]\;.
\end{equation}

{}For $G_2$, one has \cite{Parwani}

\begin{equation}
-\frac{\delta^2\lambda^2}{6} {\rm Re}[G_2 (-i \Omega,{\bf 0})] = 
H(\Omega) =
- \frac{\delta^2 \lambda^2 \Omega^2}{4 (2 \pi)^4} \int_1^\infty 
\frac{dx}{e^{\beta
\Omega x}-1}
\int_1^\infty \frac{dy}{e^{\beta \Omega y}-1} \ln \left|
\frac{\sqrt{x^2-1}+
\sqrt{y^2-1}}{\sqrt{x^2-1}-
\sqrt{y^2-1}}   \right|   
\end{equation}

\noindent
In the high temperature limit, one can  show \cite{Parwani}
that 

\begin{equation}
{}F_2 + H(\Omega) \simeq \delta^2  \frac{\lambda^2 T^2}{24 (4 \pi)^2} 
\left[
\ln \left(\frac{\Omega^2}{T^2}\right) + 5.0669 \right]\;.
\end{equation}

{}Finally, using the high temperature approximation for 
$h(y)$ and putting all 
together one gets

\begin{eqnarray}
{\rm Re}[\Sigma^{\delta^2}_5 (p)] &\simeq& 
\delta^2 \frac {\lambda^2 \Omega^2}{(32 \pi^2)^2} 
\frac {1}{\epsilon^2} +
\delta^2 \frac {\lambda^2 \Omega^2}{(32 \pi^2)^2} \frac {1}{\epsilon} + 
\delta^2 
\frac {\lambda^2 p^2}{(32 \pi^2)^2} \frac {1}{6 \epsilon} \nonumber \\
&-&\delta^2 \frac {\lambda^2}{16 \pi^2 \epsilon} \left \{ 
\frac {T^2}{24} - 
\frac {T \Omega}{ 8 \pi} 
+\frac {\Omega^2}{32 \pi^2} \left [ \ln \left (
\frac {4 \pi T^2}{\mu^2}
\right ) - \gamma_E \right ] \right \} \nonumber \\
&+&\delta^2 \frac{\lambda^2 \Omega^2}{2 (4 \pi)^4} 
\left[ \ln^2 \left( \frac{
\Omega^2}{4 \pi \mu^2} \right) + \left(2 \gamma_E - \frac{17}{6} \right) 
\ln\left(
\frac{\Omega^2}{4 \pi \mu^2} \right) + \gamma^2_E -
\frac{17 \gamma_E}{6} +
3.5140 \right ]\nonumber \\
&-& \delta^2 \frac {\lambda^2}{32 \pi^2} \left[- \ln \left ( 
\frac {\Omega^2}{4 \pi \mu^2} \right ) + 2 - \gamma_E \right ] 
\left \{  \frac {T^2}{24} - \frac {T \Omega}{ 8 \pi} 
 - \frac {\Omega^2}{16 \pi^2} \left [ \ln \left (
\frac {\Omega}{4 \pi T}\right ) +\gamma_E  - \frac{1}{2}  \right ] 
\right \} \nonumber \\
&+& \delta^2 \frac {\lambda^2 T^2}{24(4 \pi)^2}\left [ \ln \left ( 
\frac {\Omega^2}{T^2} \right ) + 5.0669 \right ]\;.
\label{div9}
\end{eqnarray}

\section{On the Renormalization at order $\delta^2$ and at order $\delta^n$}

To obtain the total finite order $\delta^2$ contribution one can add all
divergences appearing in Eqs.
(\ref{div2})-(\ref{div3}) and (\ref{div9}). 
As it can be easily seem all the
nonrenormalizable temperature dependent divergences cancel exactly and
one is left with

\begin{equation}
\Sigma^{\delta^2}_{\rm div}= \Sigma^{\delta^2}_{\rm div}(\Omega^2)+
\Sigma^{\delta^2}_{\rm div}(p^2) +\Sigma^{\delta^2}_{\rm div}(\eta^2) \; ,
\end{equation}
where
\begin{equation}
\Sigma^{\delta^2}_{\rm div}(\Omega^2)= 
\delta^2 \frac{\lambda^2 
\Omega^2}{(32 \pi^2)^2} \left 
( - \frac {2}{\epsilon^2} + \frac {1}{\epsilon} \right ) \;,
\end{equation}
\begin{equation}
\Sigma^{\delta^2}_{\rm div}(p^2)= 
\delta^2 
\frac {\lambda^2 }{(32 \pi^2)^2} \frac{p^2}{6\epsilon} \;,
\end{equation}
and
\begin{equation}
\Sigma^{\delta^2}_{\rm div}(\eta^2)= 
 \delta^2 
\frac {\lambda \eta^2}{(32 \pi^2)} \frac {1}{\epsilon} \;,
\end{equation}
By looking at all diagrams which contribute to this order one can identify two 
classes. The first is composed by diagrams such as the ones described by 
Eqs. (\ref{div4})-(\ref{div19}). All of them are 
analogous to the diagrams which appear at ${\cal O}(\lambda^2)$ in the original theory 
and can be rendered finite by similar mass and wave-function counterterms, 
which are respectively
\begin{equation}
\Sigma^{\delta^2}_{\rm ct}(\Omega^2)=B^{\delta^2}\Omega^2=- \delta^2 \frac{\lambda^2 }{(32 \pi^2)^2} \left 
( - \frac {2}{\epsilon^2} + \frac {1}{\epsilon} \right ) \Omega^2 \;,
\end{equation}
and 
\begin{equation}
\Sigma^{\delta^2}_{\rm ct}(p^2)=A^{\delta^2} p^2=-\delta^2 
\frac {\lambda^2 }{(32 \pi^2)^2} \frac{p^2}{6\epsilon} \;.
\end{equation}
The second kind of diagram is exclusive of the interpolated theory and carries 
at least one $\delta \eta^2$ vertex. At ${\cal O}(\delta^2)$ this diagram is described 
by Eq. (\ref{div2}) which displays the divergent term 
$\Sigma^{\delta^2}_{\rm div}(\eta^2)$. Looking at ${\cal L}^{\delta}_{\rm ct}$ 
one identifies a $\eta^2$ counterterm whose Feynman rule is 
$i \delta B^{\delta} \eta^2$. Since the actual pole is of order 
$\delta^2$ one then identifies the coefficient as being $B^{\delta^1}$, 
displayed in Eq. (\ref{div1}).
Then,
\begin{equation}
\Sigma^{\delta^2}_{\rm ct}(\eta^2) = - \delta B^{\delta^1} \eta^2 =
-  \delta^2 \frac{  
\lambda }{32 \pi^2 \epsilon}  \eta^2\;.
\end{equation}
In practice, the renormalization at higher orders can be done as above. 
That is, ${\cal O}(\delta^n)$ diagrams belonging to the first class will be renormalized 
exactly as in the original theory at ${\cal O}(\lambda^n)$. This is obvious from the 
fact that all the diagrams in this class are of order-$\delta^n \lambda^n$. It is easy 
to check that for those diagrams the most divergent terms will display 
$\epsilon^{-n}$ poles. On the other hand, ${\cal O}(\delta^n)$ diagrams belonging to 
the second class will make use of the counterterm $\delta B^{\delta^{n-1}} \eta^2$, 
where $B^{\delta^{1-n}}$ has been evaluated in a previous order. One can also easily 
check that for these diagrams the most divergent terms will have $\epsilon^{n-1}$ 
poles. Moreover, power counting reveals that those $\delta \eta^2$ insertions make 
the loops more convergent. For example,  {\it all} one loop diagrams of order 
${\cal O}(\delta^n)$, with $n \ge 3$ are finite. 

{}Finally, one should note that the renormalization prescription adopted here is 
in accordance with the one suggested in Ref. \cite{Hatsuda}, where the order by 
order renormalization was shown to hold at any higher orders in $\delta$.

\section{Numerical results}

One can now set $\delta=1$ and apply the PMS to the finite thermal mass. 
{}First let us set $m=0$ so that our results for the thermal mass 
can be compared directly 
with the resummed perturbative expansion (RPE) results of 
Ref.~\cite {Parwani}. 
At order-$\delta$ one gets
\begin{equation}
\bar \eta = 2\pi T \left [ \ln \left ( \frac {4\pi T^2}{\mu^2} \right ) - 
\gamma_E \right ]^{-1}\;,
\end{equation}
which, clearly, does not depend on the the coupling and cannot generate
nonperturbative information. However, nonperturbative
results appear already at second order in $\delta$. Table 1 shows our
results and the results furnished by the RPE for $M_T^2$ in units of $\mu$ for 
$\lambda=0.1$. We also show, in
units of $\mu$, the optimal values of $\eta$.

\vspace{0.5cm}

\begin{center}
\begin{tabular}{c|c|c|c}
\hline
$T/\mu$ & ${\bar \eta}/\mu$ & $O(\delta^2)$ & {\rm RPE} \\ 
\hline
0.5     & 0.033           & 0.098     &  0.099  \\
1.0     & 0.067           & 0.391     &  0.396  \\
1.5     & 0.100           & 0.880     &  0.892  \\
2.0     & 0.133           & 1.564     &  1.587  \\
2.5     & 0.166           & 2.445     &  2.481  \\
3.0     & 0.200           & 3.522     &  3.574  \\
3.5     & 0.233           & 4.794     &  4.867  \\
4.0     & 0.266           & 6.263     &  6.358  \\
4.5     & 0.299           & 7.927     &  8.049  \\
5.0     & 0.332           & 9.788     &  9.939  \\
5.5     & 0.366           &11.845     & 12.029  \\
6.0     & 0.399           &14.098     & 14.317  \\
6.5     & 0.432           &16.547     & 16.806  \\
7.0     & 0.465           &19.193     & 19.494  \\
\hline 
\end{tabular}

\vspace{0.25cm}

\centerline{ Table 1 -  Results for $M_T^2/\mu^2 \; (\times 10^{-2})$.}
\end{center}

Let us now obtain the critical temperature for the phase transition at 
${\cal O}(\delta^2)$.
Taking $\lambda=0.1$, we reset $m^2 = - \mu^2$ in $M_T^2$ observing a 
second order phase transition at  the 
critical temperature $T_c = 15.57 \; \mu$ whereas the modified 
perturbation scheme (MPS) of Banerjee and Mallik \cite {Banerjee} predicts 
$T_c = 15.63\; \mu$. Choosing $\lambda=0.01$ we find 
$T_c = 49.03\; \mu$ whereas the value  $T_c = 49.05\;\mu$ 
is predicted by the MPS. Note that in the calculation of the
critical temperature performed in Ref. \cite {Banerjee} the propagator has 
been effectively dressed 
up to the leading order correction in the temperature, which is 
set by the tadpole 
term in (\ref{delta1}) (see their Eq. (4.5)). 
Here, on the other hand, we are definitely working with all higher order
corrections up to the two-loop contribution. The fact that our results for
the critical temperature are slightly smaller than those obtained
in Ref.~\cite{Banerjee} is an indication of the importance of these higher
order corrections and is in accordance with well known results concerning
the study of phase transitions in the context of the electroweak  
effective potential beyond 1-loop  \cite{beyond}. The results are also in
accordance with recent results for the finite temperature effective potential
of the $\lambda \phi^4$ theory, obtained with the super-daisy approximation 
\cite{sato3}.   

\section{Conclusions}

Using the $\lambda \phi^4$ model we have shown how the optimized
$\delta$-expansion can be useful in extracting nonperturbative
information through an essentially perturbative evaluation of {}Feynman
graphs. Our ${\cal O}(\delta^2)$ results for the thermal mass and for
the phase transition critical temperature are in excellent agreement
with the ones given by other methods \cite{Banerjee,Parwani}. However,
although providing very similar results, these methods differ, from the
$\delta$-expansion, in some aspects which may become important if one
tries to consider higher orders. {}For example, within the latter
methods the effective mass used in the modified propagator changes order
by order turning the propagator into a coupling dependent quantity from
the start. One can then expect the selection and evaluation of higher
contributions to become complicated quickly. The $\delta$-expansion
method avoids these potential problems by using $\Omega$ in the modified
propagator which is used at any order calculation. Therefore, after
drawing the relevant graphs which contribute to a given order, one does
not have to worry about bookkeeping inconsistencies nor renormalization 
problems since this is done as in perturbation theory. 
The extension to higher order in $\delta$ is immediate and, as discussed above,
leads to a consistent resummation procedure in finite temperature field theory.

Although we have not attempted to prove the possibility of convergence of our results 
we have explicitly shown that the procedure interpolation, renormalization, 
and optimization in the finite temperature domain can be consistently handled 
to furnish encouraging results.

We also note that the linear $\delta$-expansion can be extended to the
case of gauge theories, where it has already been used as a tool to
study the electroweak phase transition on the lattice \cite{evans}.
Recently, it has been shown \cite{gromes} that the method does not spoil
gauge invariance. In this context, the linear $\delta$-expansion may be
a useful technique to analytically study the nonperturbative aspects and
difficulties associated, for example, with the electroweak phase
transition as well as other problems in high temperature gauge theories.

\acknowledgements

It is a pleasure to thank H. F. Jones and P. Parkin for their interest in 
this work
and for pointing out few mistakes in a previous version of the manuscript.
ROR was partially supported by CNPq and FAPERJ.

\newpage

\begin{center}
{\large \bf Figure Captions}
\end{center}

{\bf Figure 1:} Diagrams contributing to the self-energy 
at first order in $\delta$.

{\bf Figure 2:} Diagrams which are order $\delta^2$ contributing
to the self-energy.

\begin{figure}[b]
\epsfxsize=5cm 
{\centerline{\epsfbox{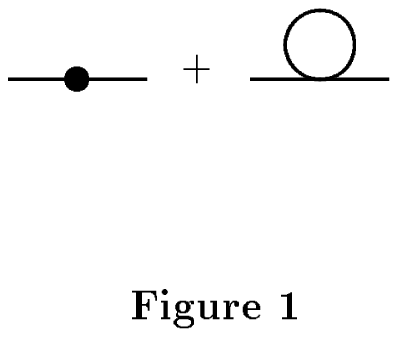}}}

\vspace{1cm}

\end{figure}


\begin{figure}[b]
\epsfxsize=15cm 
{\centerline{\epsfbox{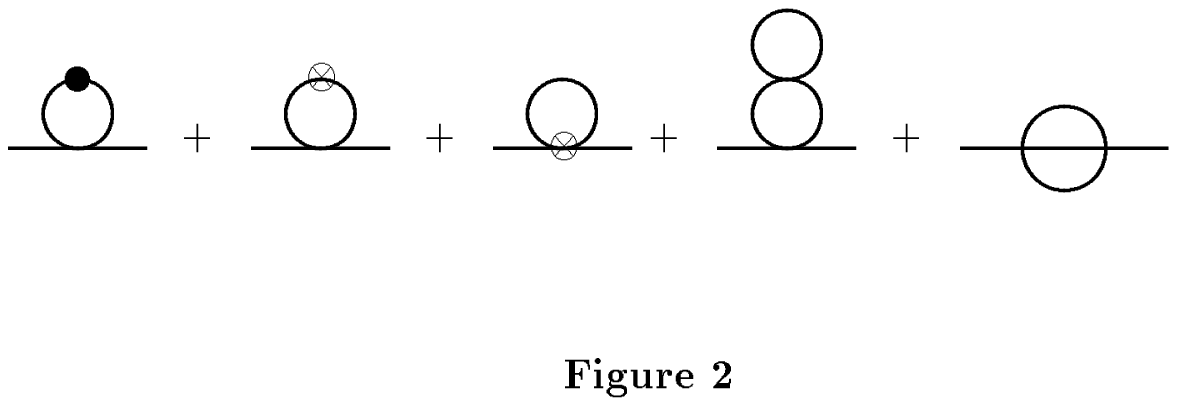}}}

\end{figure}

\end{document}